\documentclass[ twocolumn, showpacs, pra, superscriptaddress, floatfix
 amsmath,amssymb,
 aps,
floatfix,
]{revtex4-1}
\usepackage{CJK}

\usepackage{graphicx}
\usepackage{dcolumn}
\usepackage{comment}
\usepackage{xcolor}
\usepackage[colorinlistoftodos]{todonotes}
\usepackage{hyperref}
\usepackage[english]{babel}
\usepackage{bm}
\usepackage{amsmath}
\usepackage{amssymb}
\usepackage{comment}
\usepackage{subfigure}
\usepackage{epstopdf}
\usepackage{hyperref}
\usepackage{appendix}
\usepackage{float}

\usepackage{mathtools}

\DeclarePairedDelimiterX\braket[2]{\langle}{\rangle}{#1 \delimsize\vert #2}

\DeclareUnicodeCharacter{03B2}{\ensuremath{\beta}}
\UseRawInputEncoding

\makeatother

\newcommand{\Rnum}[1]{\uppercase\expandafter{\romannumeral #1\relax}}

\begin{document}


\title{Simulations and theory of power spectral density functions for time dependent and anharmonic Langevin oscillators}

\author{AbdAlGhaffar K. Amer}
 \email{amer1@purdue.edu}
 \affiliation{
 Department of Physics and Astronomy, Purdue University, West Lafayette, Indiana 47906 USA
}

\author{F. Robicheaux}%
 \email{robichf@purdue.edu}
\affiliation{
 Department of Physics and Astronomy, Purdue University, West Lafayette, Indiana 47906 USA
}

\begin{abstract}
Simulations and theory are presented for the power spectral density functions (PSDs) of particles in time dependent and anharmonic potentials including the effects of a thermal environment leading to damping and fluctuating forces. 
We investigate three one dimensional perturbations to the harmonic oscillator of which two are time dependent changes in the natural frequency of the oscillator, while the other is a time independent extension of the quadratic potential to include a quartic term. We investigate the effect of these perturbations on two PSDs of the motion that are used in experiments on trapped nano-oscillators. 
We also derive and numerically test the PSDs for the motion of a spherical nanoparticle in a Paul trap. 
We found that the simple harmonic Langevin oscillator's PSDs are good approximations for the $x$-and $y$-coordinates' PSDs for small values of the parameter $q$ of the Mathieu equation, but the difference can be more than a factor of two as '$q$' increases. 
We also numerically showed that the presence of a permanent electric dipole on the nanosphere can significantly affect the PSDs in the $x$-and $y$-coordinates.  


\end{abstract}

\maketitle

\section{\label{sec:level1}Introduction}

Nanomechanical oscillators have evolved into a wide arena with various applications in many fields. They are used as ultrasensitive sensors for charge \cite{moore2014search}, mass \cite{paul1990electromagnetic} and force \cite{monteiro2020force},\cite{geraci2010short} and in other fields like testing wave function collapse models \cite{goldwater2016testing},\cite{Barker}, physics beyond the standard model \cite{moore2021searching},\cite{aggarwal2022searching}, gravitational wave detection \cite{arvanitaki2013detecting}. The power spectral density (PSD) is a major tool in extracting information about the system. High sensitivity has been reached with narrow line-width and high quality factors. \cite{Barker} 






To function as ultrasensitive detectors, the experimental PSDs can be compared to accurate analytic expressions. Traditionally, the experimental PSDs are compared to those of a simple harmonic Langevin oscillator. However, cases with time dependent frequencies or anharmonic potential change the PSD. 
We derive analytic expressions for the PSDs and verify them numerically for several time dependent and anharmonic cases as well as a particle in a Paul trap.

We examine the case of drifting frequency as well as an oscillating frequency in time. We use numerical simulations to show that the Quadrature PSD (QPSD) of the motion, which is the PSD of the motion after removing the fast oscillations \cite{krylov1950introduction},\cite{mitropolsky1961asymptotic} and was used in \cite{Barker}, remains in agreement with the QPSD for a simple harmonic Langevin oscillator(SHLO) \cite{Langevin-oscillator}. In these cases the traditional PSD is  sensitive to the perturbations and deviates significantly from that of the SHLO PSD. We derive new analytic expressions for the PSD for these cases and show that they fit the PSDs from the numerical simulations. We also simulate the case of nonlinear oscillation \cite{hebestreit2018calibration},\cite{mitropolsky1961asymptotic},\cite{gieseler2012subkelvin} by introducing a quartic term in the potential. We show that although the PSD substantially broadens and gets blue shifted from that of the SHLO PSD, the QPSD retains a Lorentzian shape, approximating the system parameters to an accuracy of a few percent. The values of the parameters, for the simulations presented here, were chosen to roughly give the displacement frequencies of a nanoparticle in a Paul trap \cite{Barker} \cite{bullier2020characterisation}. We also explored other values of the parameters and the qualitative features for the PSD and the QPSD remain consistent. 
 
\par 


A particle in a radio frequency(RF) Paul trap has the $z$-direction in a simple harmonic motion while the $x$,$y$-directions are described by the Mathieu equations \cite{knoop2016trapped}. For certain parameter regimes, the $x$,$y$-motion has low frequency secular motion with an additional high frequency from the RF voltages \cite{Barker} \cite{bullier2020characterisation}. We derive analytic formulas for the PSDs for a particle in a Paul trap. 
The percentage difference between the Paul trap PSDs and those of the simple harmonic Langevin oscillator were independent of the friction coefficient $\Gamma$, but were heavily dependent on some trap parameters. The disagreement between the Paul trap and simple harmonic PSDs increases as the oscillating voltage of the Paul trap, $V_{RF}$, increases. We show that there are cases where the Paul trap's $x$,$y$-components PSD and QPSD differ by more than a factor of 10 from those for the SHLO.

\par
We also examine cases when the electric dipole on the nanoparticle affects the equations of motion. Our simulations showed that in such cases the power spectral densities of the $z$-component are
less sensitive to the dipole moment than for the $x$, $y$-components. For the case we investigated there was almost no change for the $z$-component's PSDs from those of a SHLO 
for small values of the parameter $q$ of the Mathieu equation. In contrast, the PSD and QPSD for the $x$,$y$-components were substantially changed when the nanoparticle had reasonable electric dipole moments.
 
\par

This paper proceeds as follows. Section \ref{sec:level2} discusses the model and the numerical technique as well as presents the main equations used. Section \ref{subsec:level1} discusses three pertubative cases for the 1-dimensional SHLO. Section \ref{subsec:level2} discusses the limitation of using the SHLO PSDs for a particle in a Paul trap. In both sections \ref{subsec:level1} and \ref{subsec:level2} the numerical simulation results are compared with the analytical expressions that are derived in Appendices \ref{sec:appendixB} and \ref{sec_app_Mathieu} while showcasing the insensitivity of the QPSD to small perturbations. Section \ref{sec:Conclusion} presents the conclusions and summarizes the results. In appendix \ref{sec:appendixB} the derivations for the analytic expressions of the PSD for the 1D perturbative cases are discussed. The derivation for the PSD and QPSD of a particle in a Paul trap is presented in appendix \ref{sec_app_Mathieu}

\section{\label{sec:level2}Methods}

The system under consideration is an oscillating classical particle in contact with a thermal environment which causes the particle to experience both damping as well as random thermal kicks. We model this system as a Langevin oscillator. The object's motion follows 
\begin{equation}
\label{eq1}
    \ddot{x} = F_x / m -\Gamma \dot{x} + F_{th} / m
\end{equation}
where $F_x$ is the trapping force of the oscillator, $\Gamma$ is the damping constant,  
$F_{th}$ is a random force representing the random collisions with the surrounding thermal environment, and $m$ is the mass of the oscillating particle. The random force $F_{th}$ is a stochastic white noise with a power spectral density of $2 k_b T m \Gamma$ \cite{kubo1966fluctuation},\cite{kubo2012statistical},\cite{landau-statistical}.

\subsection{\label{sec:level} Paul trap} 
In this section, the equations of motion of a particle in a Paul trap are given. The first section assumes no permanent electric dipole on the sphere, while the second section includes the effect of a permanent electric dipole on the sphere. 
\subsubsection{Paul trap without dipole}
\label{paul_no_di}

The potential inside the trap $V(x,y,z)$ is given by \cite{knoop2016trapped},\cite{foot2004atomic}
\begin{equation}
\label{poteq}
    V(x,y,z) = k V_{end} \frac{z^2-\frac{x^2+y^2}{2}}{z_{0}^{2}} \ - \ V_{RF} \frac{x^2-y^2}{2r_{0}^{2}} \ \cos(\Omega_{RF} t )
\end{equation}
where $V_{end}$ is the potential on the end caps in the $z$-axis, $k$ is a dimensionless constant, $z_0$ and $r_0$ are constants of length dimension, 
and $V_{RF}$ is the potential on the rods oscillating with frequency $\Omega_{RF}/2\pi$. The effects of higher order nonlinear instabilities \cite{wang1993non} aren't included in the following results because the trap potential Eq. (\ref{poteq}) is quadratic. 
The above potential produces forces on the three coordinates \{$x$, $y$ ,$z$\} of the trapped particle of the form\cite{knoop2016trapped}:
\begin{eqnarray}
    F_i/m &=& -\; \frac{\Omega_{RF}^2}{4}\left( a_i - 2 q_i \cos(\Omega_{RF} t)  \right) \ x_i(t) \label{eq7}
\end{eqnarray}
where the index $i$ runs over the three components $\{$x$,$y$,z\}$, with $-a_z/2 = a_x = a_y = - k \frac{4 Q V_{end}}{m \Omega_{RF}^{2} z_{0}^{2}}$ , $q_x = - q_y = \frac{2 Q V_{RF}}{m \Omega_{RF}^{2} r_{0}^{2}}$ and $q_z = 0$. The notation for $\{a_x, a_y, a_z\}$ is the standard notation for the parameters of the Mathieu equation and are not to be confused with the acceleration.
 The equation of motion in the $z$-component is a pure harmonic oscillator equation with natural frequency $w_z =\sqrt{a_z} \Omega_{RF}/2= \sqrt{\frac{2 Q k V_{end} }{m z_{0}^{2}}}$, while that in $x$,$y$-components follow a Mathieu equation. 
\par
The solutions of the Mathieu equation are well known.\cite{knoop2016trapped} For the cases below, we used a Floquet expansion for the stable solution with the form 
\begin{equation}
    x(t) = e^{i \omega_0 t} \; \sum_{n \in \mathbb{Z}} b_n \; e^{i n \Omega_{RF} t}
\end{equation}
Since for stable solutions, $b_{|n|} >\!> b_{|n|+1}$ for all integer $n$ \cite{knoop2016trapped}, an approximate solution for $w_0$ can be found to the desired degree of accuracy by truncating the series at a certain $\pm n_{max}$ then solving iteratively for $w_0$. This procedure gives the solution for $\omega_0$ in the form of a continued fraction.  For 
the $x$- and $y$-components
\begin{equation}
    \omega_0 = \beta \, \Omega_{RF} 
\end{equation}
with $\beta$ given by a continued fraction \cite{march2005quadrupole}, 
\begin{eqnarray}
    4 \beta^2 - a &= \left(
    \frac{q^2/4}{(1-\beta)^2 - a/4 -} \frac{q^2/4}{(2-\beta)^2 - a/4 -} \frac{q^2/4}{(3-\beta)^2 - a/4 }\right)  \nonumber \\
    &+ \left( 
    \frac{q^2/4}{(1+\beta)^2 - a/4 -} \frac{q^2/4}{(2+\beta)^2 - a/4 -} \frac{q^2/4}{(3+\beta)^2 - a/4 } \right)\nonumber \\
\end{eqnarray}

This equation was used as an iterative equation to obtain a numerical approximation for $\omega_0$. We started with  $\beta =  \frac{1}{2} \left( a_x + q_{x}^{2} /2 \right)^{1/2}$ as a first approximation \cite{march2005quadrupole} in the right hand side then the obtained value for $\beta$ was reinserted on the right hand side again to obtain a better value for $\beta$. This process was repeated until the change in $\beta$ between each two successive iterations was less than $10^{-4} \%$. 

\par

\subsubsection{Paul trap with dipole}
\label{paul_with_di}
In this section, we give the equations of motion with a nonzero permanent electric dipole $\vec{p}(t) = \{p_x(t), p_y(t), p_z(t)\}$ on the nanoparticle that has a constant magnitude but its direction changes in time. 
This will introduce an additional force on the nano-particle  which has the form $-\vec{\nabla} (- \vec{p} \cdot \vec{E})$ with $\vec{E}(x,y,z)= -\vec{\nabla} V(x,y,z)$ the electric field at the nanoparticle. This changes Eq. (\ref{eq7}) by replacing $x_i(t)$ on the right hand side with $x_i(t) + p_i(t)/Q$ for the three components $\{$x$,$y$,z\}$. 

\par
 Since the nanoparticle is spinning, the direction of the electric dipole moment vector changes over time following the differential equation \cite{goldstein2002classical}
 \begin{equation}
     \dot{\vec{p}} = \vec{\eta} \times \vec{p}
 \end{equation}
 with $\vec{\eta} = \{\eta_x, \eta_y, \eta_z\}$ where $\eta_x, \eta_y$ and $\eta_z$ are the angular frequencies of the spherical  
 nanoparticle around its center of mass. The angular frequencies also change in time due to the torque from the electric field following the equation \cite{griffiths2005introduction}:
 \begin{equation}
     \dot{\vec{\eta}} = \vec{p} \times \vec{E}/I  \label{eq18}
 \end{equation}
 where $I$ is the moment of inertial of a sphere: $\frac{2}{5} m R^2$. The damping and fluctuating terms are discussed in the next section. 
 \par


\par
Lastly, since the fourth order method that was implemented to solve the differential equation is not a symplectic method \cite{press2007numerical}, the magnitude of the electric dipole moment vector slightly changes in a time step. While the change in a time step was small, this was cumulatively the largest source of numerical error. We numerically found that renormalizing the dipole moment after each time step gave the fastest convergence to the orbit as the time step, $dt$, decreased.
\subsection{\label{subsec:level1} Numerical implementation of the thermal environment}
\label{Langevin}
To solve for the motion, we used the fourth order Runge-Kutta method \cite{press2007numerical} with an adaptive time step, RKQS in Ref. \cite{press2007numerical}. We tried other numerical solvers like the RK2 method or Euler algorithm but they were slower or were less accurate for the same step size. All the simulations have the form of Eq. \ref{eq1}. 
Both the first and the second term are in the acceleration calculator of the RKQS integrator, however the thermalization term $F_{th}$ is a stochastic term and isn't directly expressible as an acceleration. So, in order to simulate the effect of the $F_{th}$ term, we adjust the velocity $v$ after each time step $dt$ as follows.

\begin{equation}
    v \rightarrow  v + \delta v
\end{equation}
where the second term $\delta v$ is a random number that follows a Gaussian distribution. This $\delta v$ represents the effect of the random forces from the environment. These thermal kicks should lead to a Maxwell-Boltzmann distribution in the velocity at long times when the other forces are time independent. 

The thermal kick $\delta v$ is different each time step. In order to obtain the equipartition theorem, this $\delta v$ needs to follow a Gaussian distribution of the form: 
\begin{equation}
    \rho(\delta v) = \sqrt{\frac{\alpha}{\pi}} e^{-\alpha (\delta v)^2} 
\end{equation}
with 
\begin{equation}
    \alpha = \frac{m}{2 k_b T (2  \Gamma  dt)}
\end{equation}
where $m$ is the particle's mass, $k_b$ is the Maxwell Boltzmann constant, $T$ is the temperature, $\Gamma$ is the damping constant and $dt$ is the time step. This technique is similar to the ones in Ref. \cite{leimkuhler2013robust} and gives sufficiently accurate approximation to the thermalization process of the nano-particle.
\par
We implemented a similar effect for the angular frequencies. Where the three components of the frequency $\{\eta_x, \eta_y, \eta_z\}$ were adjusted after each time step with
\begin{equation}
    \eta_i \rightarrow  \eta_i + \delta \eta_i
\end{equation}
with $\delta \eta_i$ also picked from a Gaussian distribution
\begin{equation}
    \rho(\delta \eta_i) = \sqrt{\frac{\beta}{\pi}} e^{-\beta (\delta \eta_i)^2}
\end{equation}
and 
\begin{equation}
    \beta = \frac{I}{2 k_b T (2  \Gamma_{\eta}  dt)}
\end{equation}
where $\Gamma_{\eta}$ is the damping constant for the angular rotation.
\par
To decrease the effect from the boundaries in the FFT by having a finite time range $0<t<100/\Gamma$, we applied a wrapping technique as follows. The simulations were run for $0<t<111.6/\Gamma$ starting with zero velocities. Then the system was thermalized through the above procedure for $0<t<100/\Gamma$. After which the motion was let to damp out in the interval $100/\Gamma<t<111.6/\Gamma$ and the data points from this interval was added to the those in the interval $0<t<11.6/\Gamma$. We found that this wrapping procedure substantially decreased the error due to the finite time range and allowed us to do the calculation with a relatively small range.
Also, the Fourier transformation points for the Paul trap simulations were taken at intervals $dt_{FFT} < (1/4) \times 2 \pi /(\Omega_{RF})$ in order to avoid aliasing the RF oscillations. 
\subsection{Analytic expressions for the PSDs}
\subsubsection{Simple Harmonic Langevin Oscillator}
\label{methodpsd}
For a harmonic Langevin oscillator, $F_x/m = -w^2_0 x$ with $w_0$ being the natural frequency of the oscillator. The Langevin oscillators have been studied in the literature \cite{Langevin-oscillator}
and its power spectral density (PSD) as defined in Appendix \ref{apx_psd} is known to take the analytic form: 
\begin{equation}
\label{eq2}
    S_x(w) = \frac{2 \Gamma k_b T / \pi m}{(w^2 - w_0^2)^2 + \Gamma^2 w^2}
\end{equation}
This PSD has a maximum value of $\left(2k_b T \right)/\left (  \pi m \Gamma w_0^2\right )$ at $w = \pm w_0$. 
\par
In some applications it is convenient to compute the PSD relative to its average value at $w=w_0$ \cite{Barker} which we term in this paper as the Quadrature Power Spectral Density (QPSD). The steps followed to obtain it are discussed in Appendix \ref{apx_QPSD}. Near the frequency peak at $\omega=0$ for $\omega_0 \gg \Gamma$, it takes the form \cite{Langevin-oscillator}: 
\begin{equation}
\label{eq5}
    S_{R^2 R^2}(\omega) = \frac{8 \ \Gamma }{\pi (\omega^2 + \Gamma^2)} \left(\frac{k_b T}{m w_0^2}\right)^2
\end{equation}
This QPSD has a maximum value of $\frac{8}{\pi \Gamma} \left(k_b T/m w_0^2\right)^2$ at $w = 0$.

The QPSD was utilized in \cite{Barker} to remove experimental difficulties in obtaining the ideal PSD due to drift in the experimental frequency of a particle in a Paul trap. One of the aims of this paper is to quantify to what extent this QPSD would remain valid to describe the motion of a particle when its frequency of oscillation is not constant in time. 

\subsubsection{Oscillator with a linear $\omega(t)$}
This is the case for which the force $F_x $ is given by $F_x /m = -\omega^2 (t) x$  where $\omega(t)$ is given by
\begin{equation}
    \label{w(t)_lin}
    \omega(t) = w_0 \big( 1-\delta + \frac{2\delta}{\tau} t \big)
\end{equation}
In this system, the frequency of the oscillator drifts from $\omega_0(1-\delta)$ to $w_0(1 +\delta)$ over the period from $t=0$ to $t=\tau$ and $\delta \ll 1$. 
\par
We derived the PSD for this case in Appendix \ref{appendix_lin} and it approximately is given by 
\begin{equation}
    S_x(\omega) \simeq \frac{2 \Gamma k_b T}{m \pi} \frac{1}{2\delta \, \omega_0} \int_{w_0(1 - \delta)}^{\omega_0(1 +\delta)} d\bar{\omega} \, \frac{1}{\left(\omega^2-\bar{\omega}^2\right)^2+\Gamma^2 \omega^2}
\end{equation}
For this case, the QPSD didn't change from that of the simple harmonic oscillator Eq. (\ref{eq5}) as will be 
 later discussed in more detail in Section \ref{res_lin}.

\subsubsection{Oscillator with an oscillating $\omega(t)$}
This is the case for which the force $F_x $ is given by $F_x /m = -\omega^2 (t) x$  where $\omega(t)$ is given by
\begin{equation}
\label{eq_gen_osc}
     \omega(t) = \omega_0  + \Delta \omega \cos(\Omega  t)
\end{equation}
The parameters $\Delta \omega$ and $\Omega$ control the amplitude and the frequency of the oscillating part in $\omega(t)$ respectively.

For this case the particle develops side bands and oscillates at multiple frequencies $\omega_n$. These frequencies are at steps of $\Omega$ from the unperturbed motion frequency $\omega_0$, where $\omega_n = \omega_0 + n\Omega$ for all integer n.  The PSD for this case is derived in App.\ref{appendix_osc} and it takes the form: 
\begin{equation}
\label{osc_ana_psd_no_int_intext}
    S_x(\omega) \simeq \sum_n J^2_{n}\left(\frac{\Delta \omega}{\Omega}\right) \frac{\Gamma k_b T}{2 m \pi \omega^{2}_{0}} \frac{1}{(\omega - \omega_n)^2 + (\Gamma/2)^2}
\end{equation}
 where $J^2_{n}$ is the n-th order Bessel function squared. The QPSD developed minor side bands at frequencies $\pm \Omega$ as shown numerically in Section \ref{secsin}.
\subsubsection{Paul trap}
This is the case in which the oscillator's force leads to the Mathieu equation, Eq. (\ref{eq7}),  
where the constants $\{ a, q\}$ are functions of the trap parameters and determine the stability of the solutions. 
We do not know of a derivation of the PSD and QPSD for this case; they are derived in Appendix \ref{sec_app_Mathieu} 
and the PSD is 
\begin{equation}
\label{paul_an_psd_intext}
S_x(\omega)=\frac{4\left(\left\langle \bar{c}^2\right\rangle\left(\Gamma^2+4 \omega^2\right)+4 \left\langle \bar{c} \; \bar{s}\right\rangle  \Gamma \omega_0+4 \left\langle \bar{s}^2\right\rangle \omega_0^2\right)}{\pi \left(\Gamma^2+4(\omega-\omega_0)^2\right)\left(\Gamma^2+4(\omega+\omega_0)^2\right)}
\end{equation}
and the QPSD is 
\begin{equation}
\label{paul_an_QPSD_intext}
    S_{R^{2}R^{2}}= \frac{2}{\pi \Gamma} \; \frac{1}{\omega^2 + \Gamma^2} \; \left(\left\langle \bar{s}^2\right\rangle + \left\langle \bar{c}^2 \right\rangle\right)^2
\end{equation}
where $\left\langle \bar{c}^2 \right\rangle$ and $\left\langle \bar{s}^2 \right\rangle$ are the averages of the squares of the Floquet expansion coefficients of the Mathieu equation solution, Eq. (\ref{eq_cs}). 

\section{\label{sec:level3}Simulation Results}
\subsection{\label{subsec:level1} Exploring perturbations to the 1D Simple Harmonic Oscillator}
In this section, the numerical and analytical results for the PSD and QPSD are presented for multiple 1D cases 
at a temperature of 300~K. Motivated by possible experimental situations like in  \cite{bullier2020characterisation} \cite{Barker}, we analyze 
three cases where the oscillation is not purely harmonic. 
The cases we investigate are of an oscillator with a slowly oscillating frequency and that with a linearly drifting frequency in time.
For this case, the new analytic expressions for the PSD and QPSD were obtained using a WKB-like approximation in App. \ref{sec:appendixB}. We also investigate adding a quartic perturbative term to the potential. 

\par
The frequencies in all simulations are of order 100~Hz while the friction coefficient $\Gamma \leq$ 1 $s^{-1}$. Such values for the parameters were chosen to roughly give the displacement PSD of a nanoparticle inside a Paul trap, as in \cite{Barker} \cite{bullier2020characterisation}. The numerical results presented here are the average of tens of thousands of runs, each providing data for the Fourier transformation for a time interval of $100/\Gamma$. 
The numerical results were tested for the SHLO and provided an exact match with the analytic PSD and QPSD in Eqs. (\ref{eq2}) and (\ref{eq5}).

\subsubsection{Oscillator with a linear $\omega(t)$}
\label{res_lin}
\begin{figure}[tp]
    \centering
    \subfigure{\includegraphics[width=1\columnwidth]{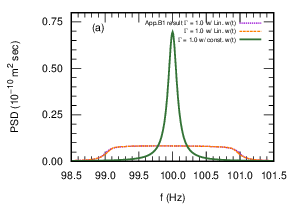} \label{fig1a}}
    \subfigure{\includegraphics[width=1\columnwidth]{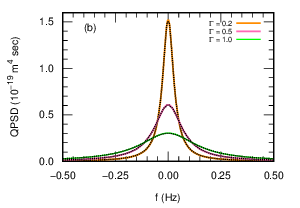} \label{fig1b}}
    
    \caption{\label{fig1} The PSD \ref{fig1a} and the QPSD \ref{fig1b} for an oscillator with a linearly changing $\omega(t)$ Eq. (\ref{w(t)_lin}). The numerical simulations' PSD agrees with the analytic expression in Eq. (\ref{eq_ana_lin_w}) obtained in App.\ref{appendix_lin} and spreads over the frequencies from 99 to 101~Hz. However the numerical simulations' QPSD remains consistent with that of a SHLO QPSD for different values of the friction coefficient $\Gamma$.}

\end{figure}

In this section, we investigate the case when the oscillation frequency changes linearly in time, Eq. (\ref{w(t)_lin}). This behavior was chosen to mimic a frequency drift in experiments which can limit the PSD linewidth, as in \cite{Barker}. For the simulations presented here, Figs. \ref{fig1a} and \ref{fig1b}, $\omega_0 = 100\times2\pi$~rad/s, $\delta =0.01$, and $\tau = 100/\Gamma$ with $\Gamma = 1$~$s^{-1}$. This represents an oscillator's frequency linearly drifting by $2\%$ over a time interval of $100/\Gamma$ around an average of $100$~Hz. 

\par

Instead of having a Lorentzian peak at $w=w_0$, the regular PSD spreads almost uniformly over the range of frequencies from 99~Hz to 101~Hz as shown in Fig. \ref{fig1a} for $\Gamma = 1$~$s^{-1}$. Such behaviour is expected because the oscillator spans all frequencies between 99~Hz to 101~Hz equally during the simulation. In Fig. \ref{fig1a}, we also plot the expression from Eq. (\ref{eq_ana_lin_w}) which can not be distinguished from the numerical simulation PSD. This spread in the PSD as well as the agreement with the expression in Eq. (\ref{eq_ana_lin_w}) was observed for values of $\Gamma$ we tested in the range [0.1 , 1.0]~$s^{-1}$ and for different small percentage drifts in $\omega(t)$ up to approximately 8\% over a period of $100/\Gamma$.
\par
The QPSD is advantageous in this case because it doesn't exhibit any major variations from the ideal formula as shown in Fig. \ref{fig1b} 
for $\Gamma =$ $0.2$,$0.5$ and $1.0$~$s^{-1}$. Our simulations showed that for small linear drifts up to approximately 8\% in the frequency over a period of $100/\Gamma$, the QPSD didn't visibly vary from the analytic result of the harmonic case in Eq. (\ref{eq5}). Therefore, the QPSD would give accurate values for the experimental parameters like the mass $m$ or the coefficient of friction $\Gamma$. 

\subsubsection{Oscillator with oscillating $\omega (t)$}
\label{secsin}
In this section we consider the situation where the frequency oscillates in time, Eq. (\ref{eq_gen_osc}). This could be a result of a slowly oscillating trapping field or general background noise that is peaking at a specific frequency as in \cite{fonseca2016nonlinear}. For our simulations $\omega_0$ is taken to be $100 \times 2\pi$~rad/sec and $\Delta \omega$ is taken to be $\xi \, \omega_0$ where $\xi$ is the fractional change in the frequency amplitude 

\begin{equation}
    \omega(t) = 100\times 2\pi  \left(1 + \xi  \ \cos(\Omega  t + \phi)\right)
    \label{eq_osc_w}
\end{equation}
where $\phi$ is a phase factor that is chosen from a flat random distribution which is different for each run.
\par
\begin{figure}[tp]
    \centering
    \subfigure{\includegraphics[width=1\columnwidth]{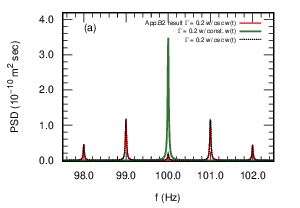} \label{fig4}}
    \subfigure{\includegraphics[width=1\columnwidth]{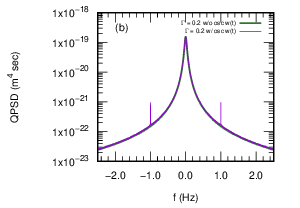} \label{fig5}}
    
    \caption{\label{fig1} The PSD \ref{fig4} and the QPSD \ref{fig5} for an oscillator with a slowly oscillating $\omega(t)$ Eq. (\ref{eq_osc_w}) with $\Omega=2\pi$~rad/s, $\xi=0.02$ and $\Gamma=0.2$~s$^{-1}$. The numerical simulations' PSD agrees with the analytic expression Appendix \ref{appendix_osc}, Eq. (\ref{osc_ana_psd_no_int}) and develops side bands at intervals of 1~Hz from the unperturbed peak at $f_0=100$~Hz. The numerical simulations' QPSD only deviates from that of a SHLO QPSD by developing two minor peaks at $\pm1$~Hz, that are 2 orders of magnitude smaller than the unperturbed peak at the origin.}

\end{figure}

As shown in Figs. \ref{fig4}, the oscillating $\omega(t)$ greatly deforms the PSD from that of a simple harmonic oscillator of Eq. (\ref{eq2}). An oscillating $\omega(t)$ with frequency $\Omega/(2\pi)$~Hz leads to the decrease in the amplitude of the regular PSD at the natural frequency of oscillation $f = f_0 = 100$~Hz and the appearance of extra peaks in the PSD at frequencies $f=f_{\pm n}$ at intervals of $\delta f = \Omega/(2\pi)$~Hz from the initial peak $f = f_0 = 100$~Hz. 
In Appendix \ref{appendix_osc}, we derived an approximate expression for the PSD, Eq. (\ref{osc_ana_psd_no_int}). To evaluate the analytic form, the summation is done for $-50<n<50$. The analytic PSD obtained from such expression is in good agreement with the numerical result as shown in Fig. \ref{fig4}. However, the approximate expression increasingly deviates from the numerical results for larger $\xi$ and $\Omega$ and for peaks further separated from $f_0$. 

\par
An interesting feature to note is that some side bands are highly suppressed in the PSD for some values of the parameters $\xi$ and $\Omega$. For instance, for $\Omega = 0.5\times2\pi$ and $\xi=0.05$ at $\Gamma =1.0$ the frequencies at $f_{\pm1} = f_0 \pm \frac{\Omega}{2\pi}$ and $f_{\pm3} = f_0 \pm 3\frac{\Omega}{2\pi}$ don't give a significant peak in the PSD. 
The cause of this behaviour is the weighting factor by the Bessel function $J_n\left(\frac{\xi \omega_0}{\Omega}\right)$. Specifically, when the fraction $\left(\frac{\xi\omega_0}{\Omega}\right)$ is close to a zero of the Bessel function $J_n(x)$, then the corresponding peak at $f_n$ becomes highly suppressed.
\par 
For the QPSD, the only change from the analytic expression in Eq. (\ref{eq5}) is the appearance of two minor peaks at frequencies of $\pm \ \Omega/(2\pi)$. In Fig. \ref{fig5}, we report the results for the case where $\Omega=2\pi$, $\xi=0.02$ and $\Gamma=0.2$.  

\subsubsection{Non-linear oscillator}
In this section, the case of having an extra non-harmonic term in the potential is considered. Anharmonicity can limit the oscillators' line-width and thus limit the accuracy of utilizing the oscillator as an accurate sensor\cite{gieseler2012subkelvin}, \cite{hebestreit2018calibration}. We consider an $x^4$ term in the potential which results if the harmonic potential was only the first order expansion of an even potential near its minimum \cite{wells2017loading}. The potential here is taken to be
\begin{equation}
    V(x) = \frac{1}{2} m w_{0}^2 x^2 \left( 1 + \alpha \ \frac{m w_{0}^2}{k T} x^2 \right) 
\end{equation}
with $\alpha$ being a dimensionless parameter giving the strength of the perturbation in units of the natural length scale of the thermal oscillator. If $\alpha = 0 $ we get the simple harmonic case, 
and as $\alpha$ increases the potential deviates more strongly from the harmonic case. 
\begin{figure}[tp]
    \centering
    \subfigure{\includegraphics[width=1\columnwidth]{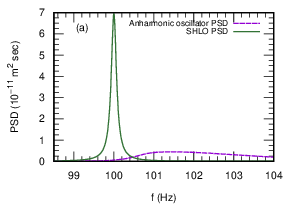} \label{fig6}}
    \subfigure{\includegraphics[width=1\columnwidth]{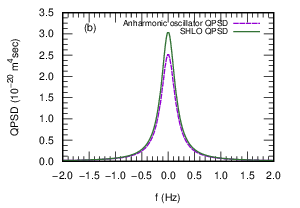} \label{fig7}}
    
    \caption{\label{fig1} The numerical simulation's PSD \ref{fig6} and QPSD \ref{fig7} for the non-harmonic potential with $\Gamma = 1$~$s^{-1}$ and the perturbation parameter $\alpha = 0.01$. The numerical simulations' PSD gets blue-shifted and widened compared to the SHLO's PSD. However the QPSD preserves the same form of the SHLO QPSD with a 15\% decrease in the peak value for $\alpha = 0.01$.}

\end{figure}
\par
As seen in Fig. \ref{fig6}, the quartic perturbation to the harmonic potential broadens the PSD and shifts it to higher frequencies. However, the QPSD approximately retains the Lorentzian distribution shape but with lower peak value as shown in Fig. \ref{fig7}. This change to the PSD and QPSD arises from the effect of the quartic perturbation on the oscillation frequency. To first order in the perturbation parameter $\alpha$, the oscillation frequency blue shifts from $\omega_0$ to $\omega_0(1+\alpha \frac{3}{2})$ \cite{bachtold2022}. This explains the blue shift in the PSD as well as the rough size of the shift, ~1.5 Hz. To explain the behaviour in the QPSD, one should notice that the oscillation frequency $\omega_0$ shows up in the denominator in Eq. (\ref{eq5}) for the QPSD. Since the quartic term blue shifts the frequency, the peak height of the QPSD will decrease. In Fig. \ref{fig7} the perturbation parameter $\alpha$ is equal to 0.01 leading to approximately a 15\% decrease in the peak value of the QPSD from that of the SHO. Our simulations showed that as $\alpha$ increases, the peak value in the QPSD decreases. However, the percentage decrease in the peak value was independent of $\Gamma$ over the range $\Gamma \in \left[0.02 , 1\right]$~$s^{-1}$.
\par
It is possible to fit the obtained QPSD to an equation of the same form as Eq. (\ref{eq5}) but with different values of the oscillation frequency $w_0$ and the friction coefficient $\Gamma$. For the case shown in Fig. \ref{fig7}, a fit of the numerically obtained QPSD gave $w_0$ approximately 1.044 times the actual value of $w_0$ and a friction coefficient, $\Gamma$, approximately 1.03 instead of 1. For many cases, the QPSD can approximate the physical parameters to an accuracy of a few percent while the PSD can no longer be fitted to the SHLO PSD at all. 






\subsection{\label{subsec:level2} Paul Trap}
 In this section, the three dimensional case of a spherical nanoparticle in a Paul trap, both with and without an electric dipole, is considered. The charge on the nanoparticle was taken to be 300 times the elementary charge and the mass was taken to be $9.6\times 10^{-17}$~kg. We used $z_0 = 3.5\times 10^{-3}$~m, $k = 0.086$, $r_0 = 1.1\times 10^{-3}$~m, $\Omega_{RF} = 5000\times 2\pi $~rad/sec., $V_{end} = 100$~V and $V_{RF} = 200$~V. These parameters were chosen to be similar to the ones in experiments on levitated nanoparticles in a Paul trap like Refs. \cite{Barker} and \cite{bullier2020characterisation}.
 \par
 
 \subsubsection{Without an electric dipole}
 \label{without_dipole}
 In this section, the motion of a particle in a Paul trap is simulated using the equation of motion Eq. (\ref{eq7}).   
 Since the motion of the $z$-component follows a SHLO equation of motion, the numerical results for the PSDs matched perfectly the analytic expressions in Eqs. (\ref{eq2}) and (\ref{eq5}). We only show the PSD and QPSD for the $x$-component because the $x$-and $y$-components follow the same equation of motion with a phase difference.   
 \par
The PSD and the QPSD for the $x$-motion are given in Eqs. (\ref{paul_an_psd}) and (\ref{paul_an_QPSD}). Here we compare those expressions and the ones for the simple harmonic Langevin oscillator(SHLO) in Eqs. (\ref{eq2}) and (\ref{eq5}) to the results we obtained from the numerical simulations. We found that for some trap parameters the SHLO PSDs are a good approximation for the Paul trap PSDs, however the differences can be substantially big for a wide range of parameters.
\par

\begin{figure}[h!] 
  \includegraphics[width=1.0\linewidth]{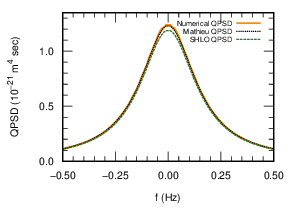} 
\caption{\label{fig9} The QPSD of the $x$-component of a particle in a Paul trap for $V_{RF} = 200$~V and $\Gamma = 1.0$~s$^{-1}$. The numerical simulation result (solid orange) is in perfect agreement with the analytic result (dotted black) Eq. (\ref{paul_an_QPSD_intext}) and deviates slightly from that of the SHLO (dashed green). The PSD results for the same parameters are discussed in the text.} 
\end{figure}

For the trap parameters above, the $a_x\sim0.0014$ and $q_x \sim0.14$ ($V_{RF}$ was multiplied by an extra geometric factor of 0.82 to match Refs. \cite{Barker} and \cite{bullier2020characterisation}). These values lead to an $x$-component PSD peak value that is approximately 1.5\% higher than that of a SHLO's PSD and a QPSD peak value that is 3\% higher than that for the SHLO's QPSD, as shown for the QPSD for the case of $\Gamma = 1.0$~$s^{-1}$ in Fig. \ref{fig9}. That percentage difference was independent of the coefficient of friction $\Gamma$, in the range $0.1<\Gamma<1$~s$^{-1}$. The numerical PSD and the QPSD were in agreement with the analytic expressions derived in Eqs. (\ref{paul_an_psd}) and (\ref{paul_an_QPSD}) respectively as shown by the dotted black curves in Fig. \ref{fig9} for the QPSD.

In fact, by examining the expressions for the PSD and the QPSD for the SHLO Eqs. (\ref{eq2}) and (\ref{eq5}) and for the Paul trap Eqs. (\ref{paul_an_psd}) and (\ref{paul_an_QPSD}), it is possible to analytically show that the percentage difference shouldn't depend on the friction coefficient $\Gamma$ nor on the temperature $T$. We will briefly go through the analysis in the next two paragraph. However, to make that deduction it is useful to first point out that all of $\left\langle \bar{c}^2 \right\rangle$ , $\left\langle \bar{c} \bar{s} \right\rangle$ and $\left\langle \bar{s}^2 \right\rangle$ are linearly proportional to $\sim \Gamma \; T$ as shown in Eq. (\ref{eq_cs}). That fact is what we used to get the Paul trap's PSD Eq. (\ref{paul_an_psd}) and QPSD Eq. (\ref{paul_an_QPSD}) dependence on both the temperature $T$ and the coefficient of friction $\Gamma$.

For the temperature dependence, both the SHLO and the Mathieu equation's PSD depend linearly on the temperature at their peak values, thus the temperature dependence cancels out in the relative difference between Eq. (\ref{eq2}) and Eq. (\ref{paul_an_psd_intext}). Similarly both the SHLO and the Mathieu equation's QPSD depend quadratically on the temperature, so the relative difference will not depend on the temperature.

The only $\Gamma$ dependence of the peak value of the SHLO QPSD Eq. (\ref{eq5}) and the Mathieu equation  QPSD Eq. (\ref{paul_an_QPSD_intext}) is a prefactor that goes like $1/\Gamma$. Thus, the $\Gamma$ dependence cancels out in the relative difference. For the PSD however, only the SHLO Eq. (\ref{eq2}) go exactly like $1/\Gamma$. The Mathieu equation PSD Eq. (\ref{paul_an_psd_intext}) only goes approximately like $1/\Gamma$ for $\Gamma \ll \omega_0$ which is the case for $0.1<\Gamma<1$.

\begin{figure}[tp]
    \centering
    \subfigure{\includegraphics[width=1\columnwidth]{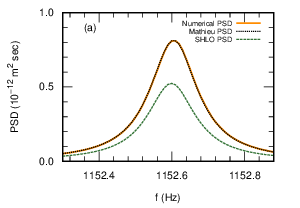} \label{fig10}}
    \subfigure{\includegraphics[width=1\columnwidth]{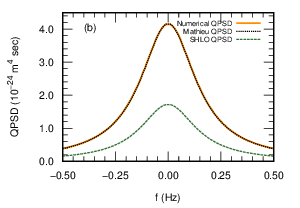} \label{fig11}}
    
    \caption{\label{fig1} The numerical simulation's PSD \ref{fig10} and QPSD \ref{fig11} for the $x$-component of a particle in a Paul trap for $V_{RF} = 870$~V, q$\sim0.6$ and $\Gamma = 1.0$~s$^{-1}$. Both the PSD and QPSD deviate from those of the SHLO while remaining in perfect agreement with the analytic expressions in Eqs. (\ref{paul_an_psd_intext}) and (\ref{paul_an_QPSD_intext}) respectively.}

\end{figure}

\begin{figure}[h!] 
  \includegraphics[width=1.0\linewidth]{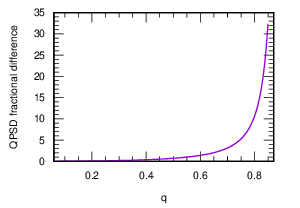} 
\caption{\label{fig_QPSD_vs_q} The fractional difference between the Mathieu QPSD and the SHLO QPSD $\big[\big(Eq. (\ref{paul_an_QPSD_intext}) - Eq. (\ref{eq5})\big) / Eq. (\ref{eq5})\big]$  increases as the parameter q increases. For the data shown here $|a_x|$ was $\sim 0.0014$ and $\Omega_{RF}/2\pi = 5\times10^3$~Hz. The range of values of the parameter $q$ considered here is inside the stable region with $0.06 \leq q \leq 0.85$.} 
\end{figure}

While the previous Paul trap parameters gave small difference from the results of a simple harmonic Langevin oscillator, the difference can be much larger for different trap parameters. For instance, in Figs. \ref{fig10} and \ref{fig11}, $V_{RF}$ is increased from 200~V to approximately 870~V leading to an increase in value of the parameter $q$ of the Mathieu equation from $q\sim0.14$ to $q\sim0.6$. This value for $q$ is still within the stable region of the Mathieu equation solution. In fact, for our case with $|a| \sim 0.0014$, the solution will still be stable for $q$ as high as $\sim 0.9$ \cite{alheit1996higher},\cite{kotana2017computation}. For such high $q\sim0.6$, the Paul trap's PSD and QPSD can no longer be approximately equated to those of a SHLO. The simulations PSD in Fig. \ref{fig10} agrees perfectly with the formula in Eq. (\ref{paul_an_psd_intext}) while its peak value is higher than that of a SHLO by approximately $55\%$. For the QPSD in Fig. \ref{fig11}, the numerical results still match the analytic formula Eq. (\ref{paul_an_QPSD_intext}) while its peak value is higher than that of the SHLO by even a higher percentage, approximately $140\%$. The difference in the peak values between the SHLO and Paul trap PSDs increases as $q$ increases, which can be seen in Fig \ref{fig_QPSD_vs_q} for $\Gamma = 1.0$~$s^{-1}$.

 \subsubsection{With an electric dipole}
 In this section, the effect of a permanent electric dipole on the sphere is included in the equations of motion. 
 If the Coulomb interaction is neglected, the $N$ charges are distributed randomly on the surface of the sphere giving a dipole roughly $\sqrt{N} \times  eR$, where $R$ is the radius of the sphere and $e$ is the charge of a proton. If the distribution is weighted by the partition function $e^{\frac{-\vec{p}.\vec{E}}{k_b T}}$ at the temperature T, then the probability of the configurations with larger dipole moments is reduced by a factor of $\sim 2$ for the parameters in our simulation. For the parameters in this section, $N=300$ leading to a rough size of $9eR$. 
 We considered the effect of electric dipoles with values ranging from $4 eR$ to $16 eR$. The values for the other parameters were taken to be the same as in the previous section, Fig. \ref{fig9}. The numerical results showed that the $x$-and $z$-components exhibit different behaviours after introducing the effect of the dipole into the equations of motion.
 \par
\begin{figure}[tp]
    \includegraphics[width=1\columnwidth]{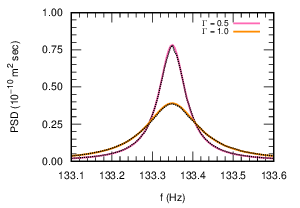}
    \caption{\label{fig14} The PSD for the $z$-component for an electric dipole of $16eR$, where $e$ is the elementary charge and $R$ is the radius of the nanosphere. The numerical PSD shown by the dotted black curves doesn't deviate from SHO PSD Eq. (\ref{eq2}) for the two value of $\Gamma = 0.5$ and $1.0$~$s^{-1}$ represented by the pink and the orange solid curves respectively.}   
\end{figure}

For the $z$-component, the PSD and the QPSD don't deviate much from the analytic formula in Eqs. (\ref{eq2}) and (\ref{eq5}). Even with the electric dipole increased to $16eR$, the numerical answer for both the QPSD and the PSD still match the analytic formulas as shown in Fig.  \ref{fig14} for the PSD with an electric dipoles of $16eR$. 
\par
\begin{figure}[tp]
    \centering
    \subfigure{\includegraphics[width=1\columnwidth]{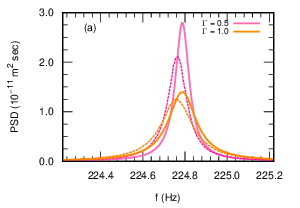} \label{figa}}
    \subfigure{\includegraphics[width=1\columnwidth]{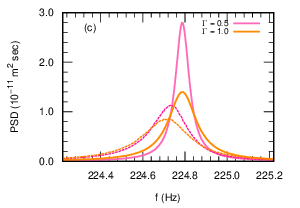} \label{figc}}
    \caption{\label{fig15} The PSD for the $x$-component for the different values of the electric dipole $p$ =$8eR$ and $16eR$ in figures \ref{figa} and \ref{figc} respectively, where $e$ is the elementary charge and $R$ is the radius of the nanosphere. The numerical PSD represented by the dashed pink  and orange curves ,for the two value of $\Gamma = 0.5$ and $1.0$~$s^{-1}$ respectively, deviates greatly as the electric dipole $p$ increases from the Paul trap's PSD given by Eq. (\ref{paul_an_psd_intext}) represented by the the pink and the orange solid curves for the two value of $\Gamma = 0.5$ and $1.0$~$s^{-1}$ respectively.} 

\end{figure}
\begin{figure}[tp]
    \includegraphics[width=1\columnwidth]{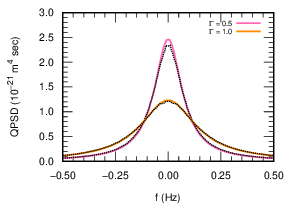}
    \caption{\label{fig13} Same as Fig. \ref{fig15}, but for the QPSD and only for an electric dipole of $16e R$. The numerical QPSD shown by the dotted black curves only deviates slightly from the Paul trap QPSD given by Eq. (\ref{paul_an_QPSD_intext}) for the two value of $\Gamma = 0.5$ and $1.0$~$s^{-1}$ represented by the the pink and the orange solid curves respectively.}   
\end{figure}

For the $x$-component, the behaviour of the QPSD is different from that of the regular PSD as we increase the electric dipole on the spherical nanoparticle. The PSD differed from Eq. (\ref{eq2}) by 23\% for an electric dipole of $8eR$ with $\Gamma=0.5$~s$^{-1}$. Also, as the electric dipole increases, the peak value of the numerical simulation's PSD decreases and the peak shifts to smaller frequencies, Fig. \ref{fig15}. We observed that the percentage decrease in the peak value from the Paul trap's PSD Eq. (\ref{paul_an_psd_intext}) is bigger the smaller the friction coefficient. From our simulations, for an electric dipole of $16eR$, when $\Gamma = 0.5$~$s^{-1}$ the decrease was about 60\% while for $\Gamma=1.0$~$s^{-1}$ the decrease was only about 40\% as shown in Fig. \ref{figc}. However, the QPSD only slightly deviates by $\approx$ 5\%  from the exact QPSD Eq. (\ref{paul_an_QPSD_intext}) as the electric dipole is increased to $16eR$ as shown in Fig. \ref{fig13}.

As the value of the parameter $q$ of the Mathieu equation is increased, the simulations PSDs deviate from the Paul trap's PSDs Eqs. (\ref{paul_an_psd_intext}) and (\ref{paul_an_QPSD_intext}) at smaller dipole moments more than for small $q$ cases. From our simulations results with the same parameters as before and for a dipole of $8eR$, $\Gamma = 1.0$, and $q\sim0.6$, the simulation's QPSD is lower by approximately $8\%$ from Eq. (\ref{paul_an_QPSD_intext}) near the peak. Also, the numerical PSD has a peak value that is approximately $40\%$ lower than the Paul trap PSD Eq. (\ref{paul_an_psd_intext}).  

\section{Conclusion}

\label{sec:Conclusion}

The numerical as well as the analytic expression for the PSD and QPSD of oscillators in several time dependent or anharmonic potentials were analyzed including the effects of a thermal environment leading to damping and fluctuating forces. 
We show here analytically and numerically that for some perturbative cases the quadrature power spectral density (QPSD) varied less than the regular power spectral density (PSD) and that it often could be approximated by the Langevin oscillator's QPSD. 

\par


We examined three possible perturbations to the harmonic oscillator potential and investigated how the PSD and the QPSD are affected by them. Two of the three cases were in the form of a slow time dependence in the oscillation frequency with the third being a time independent quartic perturbation to the potential. 
For a linear drift in the frequency of few percent over a time interval of 100/$\Gamma$, where $\Gamma$ is the damping coefficient, we found the QPSD to be unaffected \cite{Barker}, but found a significant change in the PSD from that of a Langevin harmonic oscillator, even for a drift as small as 2\% in the frequency. We presented an analytic expression for the PSD Eq. (\ref{eq_ana_lin_w}) using the WKB approximation and showed that it agrees with the PSD from the numerical simulation in Fig \ref{fig1a}. 
For a slowly oscillating frequency
the PSD was altered significantly from that of the pure harmonic oscillator, however the only change in the QPSD was the appearance of two minor side peaks. We obtained analytic expressions that matched the numerical results.
The time independent case we considered was in the form of adding a quartic term to the potential leading to an anharmonic but even potential. 
In this case 
it was observed that, even for perturbation as small as 1\% of the strength of the harmonic term, there could be a significant change in both the PSD and QPSD functions. This change stems from the fact that the perturbation blue shifts the oscillation frequency of the oscillator. 
\par
We also explored the case of a spherical particle in a Paul trap. Two situations were considered, the first one is when the electric dipole moment is neglected and the charges are assumed to be uniformly distributed on the sphere. In this case the equations of motions in the 3 coordinates decoupled and that for the $z$-component was a simple harmonic Langevin oscillator, while those for the $x$-and the $y$-components were in the form of a damped Mathieu equation with an extra Langevin-like thermalization force. 
For the $x$-and $y$-components, the agreement with the Langevin oscillator's PSD and QPSD changed with the value of the parameter $q$ of the Mathieu equation. For smaller $q$ of about $q\sim0.14$ the numerical simulation PSD and QPSD approximately matched those of a simple harmonic Langevin oscillator with only a few percent difference. However, the SHLO PSD and QPSD no longer are good approximations to the Paul trap PSD and QPSD for larger $q$ values. We derived analytic formulas for the PSD and QPSD using a Floquet expansion of the Mathieu equation solution Appendix \ref{sec_app_Mathieu}. 

We also investigated the case when there is a permanent electric dipole on the trapped nanoparticle. We considered dipoles up to $2\sqrt{N} eR$ where $N$ is the number of charges on the nanosphere and $R$ is its radius. We found that the $z$-component remained in good agreement with the Langevin oscillator's PSD Eq. (\ref{eq2}) and QPSD Eq. (\ref{eq5}) for all the electric dipole moments we examined. However for the $x$-component, the PSD and the QPSD only deviated by a few percents from the analytic formulas for small dipoles but it deviated quite significantly as the dipole moment increased to $\sqrt{N} eR /2$. The $x$-component's PSD got red shifted and its peak value decreased as the electric dipole on the nanosphere increased. The deviation from the expressions in Eqs. (\ref{paul_an_psd_intext}) and (\ref{paul_an_QPSD_intext}) showed a few percent increse as the value of the $q_x$ parameter of the Mathieu equation increased. 

Data for the figures used in this publication are available from the Purdue University Research Repository \cite{data}.



\begin{acknowledgments}
FR thanks Joseph T. Bertaux for the interesting discussions relevant to the nonlinear oscillators section. This work was supported by the Office
of Naval Research (ONR) Basic Research Challenge (BRC) under Grant No. N00014-18-1-2371.
\end{acknowledgments}

\appendix
\section{PSD and QPSD procedure}
\label{sec:appendixA}
\subsection{PSD}
\label{apx_psd}
We present here the convention we use for the PSDs of a position signal $x(t)$ over a finite time interval $\tau$. We chose
\begin{equation}
   S_x(\omega) = \lim_{\tau \to \infty} \frac{2}{\tau} |x_{\tau}(\omega)|^2 \label{psd}
\end{equation}
where $x_{\tau}(\omega)$ is the finite Fourier transform of $x(t)$ over the time interval $0<t<\tau$ and is given by
\begin{equation}
    x_{\tau}(\omega) = \frac{1}{\sqrt{2\pi}} \int_{0}^{\tau} x(t) \, e^{i \omega t} \,dt
\end{equation}

\subsection{QPSD}
\label{apx_QPSD}
The QPSD is the PSD relative to its average value at $w=w_0$. This involves doing the calculation in a frame rotating with the frequency $w_0$ \cite{Langevin-oscillator},\cite{Barker} 
. To this end the motion $x(t)$ is decomposed into two parts $x_c(t)$ and $x_s(t)$ with:
\begin{equation}
    x_c (t) + i x_s (t) = 2 x(t) e^{i \bar{\omega} t}
\end{equation}
where $\bar{\omega}$ is the peak oscillation frequency of the oscillator, $\omega_0$, or close to it.

Next, filter out the oscillations at large $\omega$ by first taking the Fourier transform of the two quadratures $x_c $ and $ x_s$ then multiplying the Fourier transforms by a filter.
The filter is chosen to ensure that the amplitude of the oscillations near $w=0$ is left unaffected while the amplitude of the oscillations at $w \sim 2w_0$ goes to zero. Then the filtered quadratures are recovered by taking the backward Fourier transform obtaining $\overline{x}_{c}(t)$ and $\overline{x}_{s}(t)$. Finally, the two quadratures are combined to get the total filtered motion as $R^2 (t) = |\overline{x}_{c}(t)|^2 + |\overline{x}_{s}(t)|^2$. The PSD of the quadrature $R^2 (t)$ is approximately the autocorrelation function \cite{Langevin-oscillator} of the total energy of the Langevin oscillator relative to its average value of $k_b T$ for the cases when $\Gamma <\!< \omega_0$ .

\section{WKB-like approximate solution for Langevin oscillator with time dependent $\omega(t)$}
\label{sec:appendixB}
Here a solution to the Langevin equation with a slowly varying $\omega(t)$ is presented. We first go through the general procedure then consider the two cases of interest, namely an oscillating and linearly varying $\omega(t)$. The equation of motion without the fluctuating force from the environment is:
\begin{equation}
    \ddot{x}  +\Gamma \dot{x} + w^{2}(t) \; x= 0 
    \label{B1}
\end{equation}


Substituting with an ansatz for $x(t) = A(t) \; e^{i \phi(t)}$ in Eq. (\ref{B1}) and setting both the imaginary and the real parts on the left hand side of the equation to zero separately yields: 
\begin{align}
2 \dot{A} \dot{\phi}+A \ddot{\phi}+\Gamma A \dot{\phi}=0 \\
\ddot{A} / A+\Gamma \dot{A} / A+\omega^2(t)-\dot{\phi}^2=0
\end{align}
The first equation gives: 
\begin{equation}
A=\frac{c}{\sqrt{\dot{\phi}}} e^{-\Gamma t / 2} \label{B4}
\end{equation}
where c is a constant of integration that is to be determined using the initial conditions. For the second equation, we apply the approximation that the higher order derivatives of $\phi$ are small compared to its first derivative $\dot{\phi}$. This WKB-type approximation is valid because we are considering cases where $\omega(t)$ is slowly varying in time. Thus, terms with $\ddot{\phi}$ are dropped compared to $\omega^2$. Then using $A$ from Eq. (\ref{B4}) one gets:
\begin{equation}
\dot{\phi} \simeq \sqrt{\omega^2(t)-\frac{\Gamma^2}{4}}
\end{equation}
Therefore, the solution under the WKB approximation is:
\begin{equation}
x_{WKB}(t)=\frac{c}{\left(\omega^2(t)-\frac{\Gamma^2}{4}\right)^{1 / 4}} e^{-\Gamma t / 2} e^{i \int_{t_0}^{t} \sqrt{\omega^2\left(t^{\prime}\right)-\frac{\Gamma^2}{4}} d t^{\prime}}
\end{equation}
\par
For the cases we consider, $\omega(t)$ is approximately a $1000\times$ larger than $\Gamma$, therefore the $\Gamma^2$ term will be dropped in the denominator and in the phase integral. Incorporating the velocity kick at time $t_l$ from the fluctuating force leads to initial conditions such that the position and the velocity are zero before $t=t_l$ and at $t=t_l$ the velocity receives a kick of $\delta v = \delta v_l$. Thus the solution from the WKB approximation $x_{WKB}(t)$ becomes:
\begin{equation}
    x_{WKB}(t) = \Theta(t-t_l) \frac{\delta v_l \; e^{-\Gamma\left(t-t_l\right) / 2}}{\left(\omega\left(t_l\right) \omega(t)\right)^{1 / 2}} \sin \left[\phi(t)-\phi\left(t_l\right)\right]  
\end{equation}
where $\Theta(t)$ is the Heaviside step function and
\begin{eqnarray}
    \phi(t) - \phi(t_l) = \int_{t_l}^t \omega (t')  \, dt'
    \label{eq_phi}
\end{eqnarray}
Since the system receives multiple thermal kicks, each at different $t_l$, the full time dependent motion is the sum of solutions each starting at a different time $t_l$. Thus the trajectory takes the form:
\begin{equation}
\label{b9}
x(t)=\sum_{l} \Theta\left(t-t_l\right)\left(\frac{\delta v_l \;    e^{-\Gamma\left(t-t_l\right) / 2}}{\sqrt{\omega\left(t_l\right) \omega(t)}} \right) \sin \left[\phi(t)-\phi\left(t_l\right)\right] 
\end{equation}
In the next two subsections we will use this solution to obtain the approximate PSD for the cases of a slowly oscillating $\omega(t)$ and a linearly varying $\omega(t)$.

\subsection{PSD for a linearly drifting $\omega(t)$}
\label{appendix_lin}
This is the case for which $\omega(t)$ is taken to be
\begin{equation}
    \omega(t) = w_0 \big( 1-\delta + \frac{2\delta}{\tau} t \big)
\end{equation}
where the frequency drifts from $\omega_0(1-\delta)$ to $w_0(1 +\delta)$ over the period from $t=0$ to $t=\tau$ and $\delta <\,< 1$. 

The exponentially decaying term in Eq. (\ref{b9}) guaranties that for each term in the sum, $(t-t_l)$ is of order $\frac{1}{\Gamma} \ll \tau$. Thus, the $\omega(t)$ in the denominator can be approximated to $\omega(t_l)$. In the difference 
\begin{equation}
    \phi(t) - \phi(t_l) = \omega(t_l) (t-t_l) + \tfrac{1}{2} \dot{\omega}(t_l) (t - t_l)^2
\end{equation}
Since we are considering a perturbative case with $\dot{\omega}(t_l)\times(t-t_l)\ll\omega(t_l)$, the first term will be the dominant one in the expansion. So we can substitute $\omega(t_l) (t-t_l)$ for $\phi(t) - \phi(t_l)$ in Eq. (\ref{b9}). Thus, $x(t)$ becomes
 \begin{equation} 
x(t) \cong \sum_{l} \Theta\left(t-t_l\right)\left(\frac{\delta v_l \; e^{-\Gamma\left(t-t_l\right) / 2}}{\omega(t_l)} \right) 
\sin \left[\omega(t_l) (t-t_l)\right] 
\end{equation}
For which the Fourier transform is 
\begin{equation}
x(\omega) \cong \sum_l \frac{\delta v_l}{\sqrt{2 \pi}} e^{i \omega t_l}\left[\frac{1}{\omega(t_l)^2-(\omega+i \Gamma / 2)^2}\right]
\end{equation}
The statistical average for  $\left\langle\delta v_l \delta v_{l^{\prime}}\right\rangle=\delta_{l l^{\prime}} \frac{2 \Gamma k_B T}{m} \delta t_l$ is used to obtain the PSD 
\begin{eqnarray}
\label{eq_ana_lin_w}
    S_x(\omega) &\simeq&  \sum_l \frac{2 \Gamma k_b T}{m \pi \tau} \frac{\delta t_l}{\left(\omega^2-\omega(t_l)^2\right)^2+\Gamma^2 \omega^2} \nonumber \\
    &\simeq& \frac{2 \Gamma k_b T}{m \pi} \frac{1}{2\delta \, \omega_0} \int_{w_0(1 - \delta)}^{\omega_0(1 +\delta)} d\bar{\omega} \, \frac{1}{\left(\omega^2-\bar{\omega}^2\right)^2+\Gamma^2 \omega^2} \nonumber \\
\end{eqnarray}
This expression gives an approximately $\Gamma$ independent flat PSD between $\omega_0(1-\delta)$ and $\omega_0(1+\delta)$ shown in Fig. \ref{fig1a}.

\subsection{PSD for a slowly oscillating $\omega(t)$}
\label{appendix_osc}
For this case we have
\begin{equation}
    \omega(t) = \omega_0  + \Delta \omega \cos(\Omega  (t-t_l) + \xi_l )
\end{equation}
where $\xi_l = \Omega t_l$ and was intruduced to simplify the expression for $\phi(t) - \phi(t_l)$ in Eq. (\ref{b9}) 
\begin{equation}
    \phi(t) - \phi(t_l) = \omega_0 (t-t_l) + \frac{\Delta \omega}{\Omega} \sin(\Omega (t-t_l) + \xi_l) - \frac{\Delta \omega}{\Omega} \sin(\xi_l)
\end{equation}


\par
Simplifying $\sin \left[\phi(t)-\phi\left(t_l\right)\right]$ in Eq. (\ref{b9}) using $\sin(a - b) = \sin(a) \cos(b) - \cos(a) \sin(b)$,
 the terms that are functions of time can be written as sums of Bessel functions $J_n\left(\frac{\Delta \omega}{\Omega}\right)$ using the relations:
\begin{equation}
    \begin{aligned}
    \cos \left( \eta \sin(\beta) \right) =& J_0\left(\eta\right) + 2  \sum_{k=1}^{\infty} J_{2 k}\left(\eta\right) \cos \left(2 k \beta\right) \\
    \sin \left( \eta \sin(\beta) \right) =&
    2 \sum_{k=1}^{\infty} J_{2 k-1}\left(\eta\right) \sin\left((2 k-1) \beta\right)
    \end{aligned}
\end{equation}
with $\eta = \Delta\omega/\Omega$ and $\beta = \Omega t$. 
This shows that the motion $x(t)$ will have several oscillation frequencies at intervals of $\Omega$ from the initial frequency $\omega_0$, in agreement with our numerical result in Sec.\ref{secsin}, such that:
\begin{equation}
    x(t) = \sum_{n=-\infty}^{\infty} \bar{x}_n(t)
\end{equation}
where $\bar{x}_n(t)$ are given by:
\begin{equation}
\begin{aligned}
\label{b19}
    \bar{x}_n(t) = \ &a_n \ J_{|n|}\left(\frac{\Delta \omega}{\Omega}\right) \sum_{l} \Theta\left(t-t_l\right) \; \left(\frac{\delta v_l \;    e^{-\Gamma\left(t-t_l\right) / 2}}{\sqrt{w(t) w(t_l)}} \right) \\
    &\quad   \ \times \ \sin\big(\bar{w}_n (t-t_l) + n \xi_l - \frac{\Delta\omega}{\Omega} \sin(\xi_l) + n\frac{\pi}{2}\big)
    \end{aligned}
\end{equation}
with $a_n = (-1)^{n/2}$ for even $n$ and $a_n = (-1)^{(n-1)/2}$ for odd n. $\bar{x}_n(t)$ is the part of the motion that is oscillating at frequency $\pm \bar{w}_n$. 
It is also approximately the same as that of a perfect Langevin oscillator with a constant natural frequency of $\bar{w}_n$ instead of $w_0$ and an extra phase. 

An approximate expression for the PSD of $x(t)$ can be found by approximating both $\omega(t)$ and $w(t_l)$ in the denominator in Eq. (\ref{b19}) to $w_0$. This approximation is valid because the maximum change in $\omega(t)$ is of order $\Delta \omega$ and we consider perturbative cases where $\frac{\Delta \omega}{\omega_0} \ll 1$. 

The Fourier transform of $x(t)$ can be evaluating by summing the Fourier transforms of $\bar{x}_n(t)$, which is given by:
\begin{equation}
\begin{aligned}
\bar{x}_n(\omega)= &J_{n}\left(\frac{\Delta \omega}{\Omega}\right)\sum_l \frac{\delta v_l}{\sqrt{2 \pi}} \frac{1}{w_0} e^{i \omega t_l} \int_{t_l}^\tau e^{-\frac{\Gamma}{2}\left(t-t_l\right)} e^{i \omega\left(t-t_l\right)}\\
& \times \sin \left[\bar{\omega}_n\left(t-t_l\right) + n \xi_l- \frac{\Delta\omega}{\Omega} \sin(\xi_l) + n\frac{\pi}{2}\right]  d t
\end{aligned}
\end{equation}
 Assuming that the upper limit is such that $\Gamma\left(\tau-t_l\right)\gg1$, this integral evaluates to:
\begin{equation}
\begin{aligned}
\bar{x}_n(\omega) \cong \ &J_{n}\left(\frac{\Delta \omega}{\Omega}\right) \sum_l \frac{\delta v_l}{\sqrt{2 \pi}} \frac{e^{i \frac{\omega}{\Omega} \xi_l}}{\omega_0} \frac{a_n}{2} \ \times
\\&\left[\frac{e^{-i \frac{\Delta\omega}{\Omega} \sin{\xi_l}}e^{i \frac{n \pi}{2}} e^{i n \xi_l}}{\omega+\bar{\omega}_n+i \Gamma/2}-\frac{e^{i \frac{\Delta\omega}{\Omega} \sin{\xi_l}}e^{-i \frac{n \pi}{2}} e^{-i n \xi_l}}{\omega-\bar{\omega}_n +i\Gamma/2}\right]
\end{aligned}
\end{equation}
The summation over $l$ is evaluated by applying the statistical averaging of the thermal kicks $\delta v_l$
\begin{equation}
\label{eq_deltav_avg}
    \left\langle\delta v_l \delta v_{l^{\prime}}\right\rangle=\delta_{l l^{\prime}} \frac{2 \Gamma k_B T}{m} \; \delta t_l
\end{equation}
where $\delta t_l$ is the time interval between the thermal kick $\delta v_l$ and the kick before it and is related to $\xi_l$ through $\delta \xi_l = \Omega \delta t_l$. We numerically tested that a good approximation for the PSD can be obtained by ignoring the interference terms between the oscillations at different $\omega_n$. Thus, the PSD becomes a sum of Lorentzians of the form:
\begin{equation}
\label{osc_ana_psd_no_int}
    S_x(\omega) \simeq \sum_n J^2_{n}\left(\frac{\Delta \omega}{\Omega}\right) \frac{\Gamma k_b T}{2 m \pi \omega^{2}_{0}} \frac{1}{(\omega - \omega_n)^2 + (\Gamma/2)^2}
\end{equation}
This is the analytic form that was compared against the numerical simulations' results, with the summation in Eq. (\ref{osc_ana_psd_no_int}) evaluated between $n=-50$ and $n=50$.

\section{PSD and QPSD using the Floquet expansion for the Mathieu equation's solution}
\label{sec_app_Mathieu}
The equation of motion for both the $x$-and $y$-coordinates of a particle in a Paul trap are damped Mathieu equations of the form
\begin{equation}
    \ddot{x}(t) = -\; \frac{\Omega^2}{4}\big( a' - 2 q' \; \cos(\Omega t)  \big) \ x(t) - \Gamma \dot{x}(t)
\end{equation}
with $\Omega$ the radio frequency of the trap and the constants \{$a'$,$q'$\} are parameters of the trap dimensions and strength. This equation can be converted to an undamped Mathieu equation by using a change of variables $x(t) = e^{-\Gamma t/2} u(t)$ \cite{zhao2002parametric}.
\begin{equation}
    \ddot{u}(t) = -\; \frac{\Omega^2}{4}\big( a - 2 q \; \cos(\Omega t)  \big) \ u(t)
\end{equation}
with the definition:
\begin{equation}
    a = a'-\frac{\Gamma^2}{\Omega^2}  \quad , \quad q = q'
\end{equation}

The solutions of the Mathieu equations are known\cite{knoop2016trapped}. For the analysis here, we did a Floquet\cite{floquet1883equations} expansion for the stable solution with the form 
\begin{equation}
    u(t) = e^{i \omega_0 t} \; \sum_{n \in \mathbb{Z}} b_n \; e^{i n \Omega t}
\end{equation}
where an iterative equation for $\omega_0$ is obtained by substituting with the above expansion back in the differential equation giving:
\begin{equation}
    \omega_0 = \beta \, \Omega 
\end{equation}
with $\beta$ given by a continued fraction \cite{march2005quadrupole}, 
\begin{eqnarray}
    4 \beta^2 - a &= \left(
    \frac{q^2/4}{(1-\beta)^2 - a/4 -} \frac{q^2/4}{(2-\beta)^2 - a/4 -} \frac{q^2/4}{(3-\beta)^2 - a/4 -} \dotsi\right)  \nonumber \\
    &+ \left( 
    \frac{q^2/4}{(1+\beta)^2 - a/4 -} \frac{q^2/4}{(2+\beta)^2 - a/4 -} \frac{q^2/4}{(3+\beta)^2 - a/4 -} \dotsi\right)\nonumber \\
\end{eqnarray}
Note that the difference in the form of the continued fraction here from that in \cite{march2005quadrupole} comes from the different convention for the form of the Mathieu equation. To obtain the solution coefficients $\{b_n\}$, we need to use the initial conditions at the time $t_l$, when the trapped particle receives a thermal kick in the velocity of value $\delta v_l$. These initial conditions have the form
\begin{eqnarray}
    x(t_l) = u(t_l) &=& 0 \label{u0}\\
    \dot{x}(t_l) = \dot{u}(t_l) &=& \delta v_l \label{dotu0}
\end{eqnarray}
     
To this end, it is convenient to rewrite the Mathieu equation in the following form:
\begin{equation}
\label{eq_Mathieu_ul}
    \ddot{u}(t) = -\; \frac{\Omega^2}{4}\left( a - 2 q \; \cos(\Omega (t-t_l) + \phi_l  \right) \ u(t)
\end{equation}
where $\phi_l = \Omega \; t_l$ and the general solution becomes 
\begin{equation}
\begin{aligned}
\label{eq_ul}
u_l(t) = \; \Theta(t-t_l)& \;
\Big\{ s_l  \sum_{n \in \mathbb{Z}} \alpha_n \sin\big((\omega_0 + n \Omega)(t-t_l) + n \phi_l\big) \\
+ &c_l  \sum_{n \in \mathbb{Z}} \alpha_n \cos\big((\omega_0 + n \Omega)(t-t_l) + n \phi_l\big) \Big\}
\end{aligned}
\end{equation}
where the solution was expanded in terms of $\sin$ and $\cos$ instead of exponentials, and $\Theta\left(t-t_l\right)$ is the step function that is zero before $t=t_l$ and equal to one for t larger than $t_l$. The $\alpha$'s are obtained by substituting the solution Eq. (\ref{eq_ul}) into the differential equation Eq. (\ref{eq_Mathieu_ul}) and equating the coefficients of each frequency to zero. This gives the recurrence relation\cite{acar2016floquet}
\begin{equation}
\label{eq_alpha_n_rec}
    \alpha_n =- \frac{q(\alpha_{n+1} + \alpha_{n-1})}{4(n + \frac{\omega_0}{\Omega})^2 - a}  \qquad , \; \forall \; \; n \in \mathbb{Z} \; \& \; n \neq 0
\end{equation}
Since the overall coefficients $c_l$ and $s_l$ were pulled outside the sum, we can choose $\alpha_0 = 1$ without loss of generality. Also because the $\{\alpha_n\}$ series is a decreasing series, it is traditionally truncated at some $\pm n_{max}$. For our numerical calculations we took $n_{max}=10$. And the $\alpha$'s are calculated iteratively by substituting in the recurrence relation Eq. (\ref{eq_alpha_n_rec}) starting with $\alpha_0 = 1$ and all other $\{\alpha_n\} = 0$ for $1<n<n_{max}$. We made sure that the change in any $\{\alpha_n\}$ between the last 2 iterations was less than 0.1\%.
\par
The constants $\{s_l,c_l\}$ were obtained by substituting in Eqs.(\ref{u0}) and (\ref{dotu0}) for the initial conditions giving

\begin{equation}
\begin{aligned}
    s_l &= \; \; \; \delta v_l \; \frac{\sigma_2}{\sigma_1 \sigma_4 + \sigma_2 \sigma_3} \\
    c_l &= -\delta v_l \; \frac{\sigma_1}{\sigma_1 \sigma_4 + \sigma_2 \sigma_3}
\end{aligned}
\end{equation}
where
\begin{equation}
\label{eq_sigmas}
    \begin{aligned}
     \sigma_1 =& \sum_{n \in \mathbb{Z}} \alpha_n  \sin(n\phi_l) \quad \quad \sigma_3 =& \sum_{n \in \mathbb{Z}} \alpha_n (\omega_0 + n\Omega) \cos(n\phi_l) \\
    \sigma_2 =& \sum_{n \in \mathbb{Z}} 
    \alpha_n  \cos(n\phi_l) \quad \quad \sigma_4 =& \sum_{n \in \mathbb{Z}} \alpha_n (\omega_0 + n\Omega) \sin(n\phi_l)
    \end{aligned}
\end{equation}

Finally, a particle in a Paul trap receives multiple thermal kicks over the course of its trajectory. Thus, the full trajectory will be a sum of multiple solutions of the form Eq. (\ref{eq_ul}), each starting at a different initial time $t_l$. Therefore, the full trajectory $x(t)$ of a particle in a Paul trap will be
\begin{equation}
\begin{aligned}    
    x(t) = \sum_l  \; &\Theta\left(t-t_l\right) e^{-\Gamma\left(t-t_l\right) / 2} \; \times \\
    \Big\{ &s_l  \sum_{n \in \mathbb{Z}} \alpha_n \sin\big((\omega_0 + n \Omega)(t-t_l) + n \phi_l\big) \\
+ &c_l  \sum_{n \in \mathbb{Z}} \alpha_n \cos\big((\omega_0 + n \Omega)(t-t_l) + n \phi_l\big) \Big\}
\label{u(t)_final}
\end{aligned}
\end{equation}
where $t_l$ takes all values from zero to the final time of the run $\tau$ in steps of $\delta t$. In the next two subsections, we will derive the PSD and the QPSD for this solution.
\subsection{PSD}
It is straightforward to get the PSD for Eq. (\ref{u(t)_final}). We just substitute with its Fourier transform in Eq. (\ref{psd}). However, since the interest is in the peaks near $\omega_0$, we only include the terms oscillating at such frequency when taking the Fourier transform. 

\begin{alignat}{1}
    S_x(\omega)\simeq \lim_{\tau \to \infty} \frac{2}{\tau} \Bigg| \sum_l \frac{1}{\sqrt{2 \pi}} \int_0^{\tau} e^{i \omega t} &e^{-\frac{\Gamma}{2} t}\Big(s_l \sin (\omega_0 t) \nonumber \\
    + &c_l \cos (\omega_0 t)\Big) \; dt \; \Bigg|^{2} \nonumber  
\end{alignat}
\begin{alignat}{1}
    \qquad \simeq \lim_{\tau \to \infty} \frac{2}{\tau}  \Bigg| \sum_l -\frac{1}{\sqrt{2 \pi}} \Bigg[&\frac{c_l-i s_l}{2} \frac{1}{i(\omega_0+\omega)-\frac{\Gamma}{2}} \nonumber\\
    +&\frac{c_l+i s_l}{2} \frac{1}{i(\omega-\omega_0)-\frac{\Gamma}{2}}\Bigg] \Bigg|^2
\end{alignat}

The method used to evaluate this expression is to substitute by the phase average and statistical average values for $c_l$ and $s_l$ inside the sum. The statistical average of $c_l$ and $s_l$ is obtained through their linear dependence on $\delta v_l$, which has a statistical average given by Eq. (\ref{eq_deltav_avg}). The phase average of $c_l$ and $s_l$ was evaluate by numerically integrating over $\phi_l$ from 0 to $2\pi$ then dividing the result by $2 \pi$. After taking the statistical average and the phase average, the terms inside the sum become independent of $l$. Thus, we obtain a sum over $\delta t$ which gives $\tau$ that cancels the $\tau$ in the denominator. We will use $\left\langle \bar{s}^2\right\rangle$,  $\left\langle \bar{c}^2\right\rangle$ and $\left\langle \bar{c} \bar{s} \right\rangle$ for the average values of $s_l s_{l'}$, $c_l c_{l'}$ and $c_l s_{l'}$ divided by $\delta t$ respectively. And thus the PSD is:

\begin{equation}
\label{paul_an_psd}
S_x(\omega)=\frac{4\left(\left\langle \bar{c}^2\right\rangle\left(\Gamma^2+4 \omega^2\right)+4 \left\langle \bar{c} \; \bar{s}\right\rangle  \Gamma \omega_0+4 \left\langle \bar{s}^2\right\rangle \omega_0^2\right)}{\pi \left(\Gamma^2+4(\omega-\omega_0)^2\right)\left(\Gamma^2+4(\omega+\omega_0)^2\right)}
\end{equation}
where 
\begin{equation}
    \begin{aligned}
    \label{eq_cs}
    \langle\bar{c}^2\rangle &=& \frac{2 \Gamma k_B T}{m} \frac{1}{2\pi}\int_0^{2\pi} \bigg(\frac{\sigma_1}{\sigma_1 \sigma_4 + \sigma_2 \sigma_3}\bigg)^2 d\phi_l  \\
    \langle\bar{c}\bar{s}\rangle &=&- \frac{2 \Gamma k_B T}{m} \frac{1}{2\pi}\int_0^{2\pi} \frac{\sigma_1 \sigma_2}{(\sigma_1 \sigma_4 + \sigma_2 \sigma_3)^2} d\phi_l  \\
    \langle\bar{s}^2\rangle &=& \frac{2 \Gamma k_B T}{m} \frac{1}{2\pi}\int_0^{2\pi} \bigg(\frac{\sigma_2}{\sigma_1 \sigma_4 + \sigma_2 \sigma_3}\bigg)^2 d\phi_l  
    \end{aligned}
\end{equation}
with $\sigma_1$, $\sigma_2$, $\sigma_3$ and $\sigma_4$ given by Eq. (\ref{eq_sigmas}). 
\subsection{QPSD}
For the QPSD we will follow the same steps discussed in Sec.\ref{methodpsd}. First, $x_c(t)$ and $x_s(t)$ are calculated from the term in $x(t)$ that is oscillating at a frequency $w_0$. This gives:
\begin{equation}
\begin{aligned}
 x_c(t)=&\sum_l \Big( s_l \sin(\psi_l(t)) + c_l \cos(\psi_l(t)) \Big)  e^{-\Gamma\left(t-t_l\right) / 2} \Theta\left(t-t_l\right) \\
x_s(t)=&\sum_l \Big( s_l \cos(\psi_l(t)) - c_l \sin(\psi_l(t)) \Big)  e^{-\Gamma\left(t-t_l\right) / 2} \Theta\left(t-t_l\right)
\end{aligned}
\end{equation}
where $\psi_l(t) = (\omega_0-\bar{\omega})(t - t_l) - \bar{\omega} t_l$. From which one obtains $R^2(t)$ to be:
\begin{equation}
\begin{aligned}
 R^2(t)=e^{-\Gamma t} \sum_{l,l^{\prime}} &\big[\left(s_l s_{l^{\prime}}+c_l c_{l^{\prime}}\right) \cos \left(w_0\left(t_l-t_{l^{\prime}}\right)\right) \\
& -\left(s_l c_{l^{\prime}}-c_l s_{l^{\prime}}\right) \sin \left(\omega_0\left(t_l-t_{l^{\prime}}\right)\right)\big] \\
&e^{\Gamma\left(t_l+t_{l^{\prime}}\right)/2} \Theta\left(t-t_{\rangle}\right)
\end{aligned}
\end{equation}
where $t_{\rangle}$ is the larger between $\{t_l , t_{l^{\prime}}\}$. The power spectral density of $R^2(t)$ is found by substituting in Eq. (\ref{psd}) 
\begin{equation}
\begin{aligned}
 S_{R^{2}R^{2}}=&\frac{1}{\pi \tau} \frac{1}{\omega^2+\Gamma^2} \sum_{l, l^{\prime}, l^{\prime \prime}, l^{\prime \prime \prime}} e^{-\Gamma / 2 \left(\left|t_l-t_{l^{\prime}}\right|+\mid t_{l^{\prime \prime}}-t_{l^{\prime \prime \prime}} \mid)\right.} \\
& e^{i \omega\left(t_l+t_{l^{\prime}}-t_{l^{\prime \prime}}-t_{l^{\prime \prime \prime}}\right) / 2} \; e^{i \omega\left(\left|t_l-t_{l^{\prime}}\right|-\left|t_{l^{\prime \prime}}-t_{l^{\prime \prime \prime}}\right|\right) / 2} \\
&\left[\left(s_l s_{l^{\prime}} + c_l c_{l^{\prime}}\right)\cos_{l,l^{\prime}} + \left(s_l c_{l^{\prime}} - c_l s_{l^{\prime}}\right)\sin_{l^{\prime},l}\right] \\
\times &\left[\left(s_{l^{\prime \prime}} s_{l^{\prime \prime \prime}} + c_{l^{\prime \prime}} c_{l^{\prime \prime \prime}}\right)\cos_{l^{\prime \prime},l^{\prime \prime \prime}}\right. \\
&+ \left.\left(s_{l^{\prime \prime}} c_{l^{\prime \prime \prime}} - c_{l^{\prime \prime}} s_{l^{\prime \prime \prime}}\right)\sin_{l^{\prime \prime \prime},l^{\prime \prime}}\right] 
\end{aligned}
\end{equation}
Where $\cos_{l, l^{\prime}}$ was used for $\cos\left[\omega_0\left(t_{l}-t_{l^{\prime}}\right)\right]$ to simplify the expression. Using the statistical averaging process, this 4-sum would give 4 possible contraction over the summation indices. The non zero ones are when $l=l^{\prime \prime}$ and $l^{\prime}=l^{\prime \prime \prime}$ or when $l=l^{\prime \prime \prime}$ and $ l^{\prime}=l^{\prime \prime}$. After simplification we obtain:
\begin{eqnarray}
\label{paul_an_QPSD}
    S_{R^{2}R^{2}}&=& \lim_{\tau \to \infty} \frac{2}{\pi \Gamma} \; \frac{1}{\omega^2 + \Gamma^2} \; \left(\left\langle \bar{s}^2\right\rangle + \left\langle \bar{c}^2 \right\rangle\right)^2 (1 - \frac{1}{\tau \Gamma}+\frac{1}{\tau \Gamma} e^{-\Gamma \tau}) \nonumber\\ 
    &=&\frac{2}{\pi \Gamma} \; \frac{1}{\omega^2 + \Gamma^2} \; \left(\left\langle \bar{s}^2\right\rangle + \left\langle \bar{c}^2 \right\rangle\right)^2
\end{eqnarray}
with $\langle \bar{s}^2\rangle$ and $\langle \bar{c}^2\rangle$ the same as defined in the previous section.

\newpage 
\bibliographystyle{unsrt} 
\bibliography{bibliography} 

\end{document}